\documentclass[epj]{svjour}

\usepackage{graphicx}
\usepackage{fancyhdr}
\def\makeheadbox{{%  remove EPJC header
 }}
\def\mytitle{My title} 
\def\myauthors{My name}  
\def\mytype{My type of session}
\def\mysession{My session}
\def\mytitle{Discovery Potential for the SM Higgs Boson in the
  Inclusive Search Channels} %Put your title here!
\def\myauthors{Alexander Schmidt}    %Put your name here!
\def\mytype{Contributed Talk}    
\def\mysession{Colliders - Higgs Phenomenology}
\begin{document}
\title{Discovery Potential for the SM Higgs Boson in the
  Inclusive Search Channels}
\author{Alexander Schmidt
%\thanks{\emph{Email:Alexander.Schmidt@cern.ch} Insert  Email  of corresponding author here}%
% \and
% Second author\inst{2}% etc
}                     % Do not remove
\institute{Institut f\"ur Experimentelle Kernphysik, Universit\"at Karlsruhe, now at Physik-Institut, Universit\"at Z\"urich, On Behalf of the ATLAS and CMS Collaborations}
%\and the second institute
%address here}
\date{}
\abstract{
This paper gives an overview of the potential to discover a Standard Model Higgs Boson
in the inclusive search channels at the ATLAS and CMS experiments at the LHC. The
most important decay modes, $\rm H \to \gamma \gamma$, $\rm H
\to WW \to ll\nu\nu$ and $\rm H \to ZZ \to 4l$ are described and a
summary of  recently published analyses using realistic
detector simulations is presented.
\PACS{
      {14.80.Bn}{Standard-model Higgs bosons}
%      {PACS-key}{discribing text of that key}
     } % end of PACS codes
}%end of abstract
\maketitle
\section{Introduction}
\label{intro}
The allowed decay modes of the Higgs Boson are predicted within the Standard
Model and depend only on its mass $m_H$. Direct searches conducted at
LEP  have given a lower limit  of $m_H >
114.4$~GeV$/c^2$ at the 95\% confidence limit~\cite{SMHiggsAtLEP}. In
the low mass region, $m_H < 150$~GeV$/c^2$ the small width of the Higgs Boson $\Gamma_H <
1$~GeV$/c^2$ can be utilized to find a narrow peak in the $\rm H \to
\gamma\gamma$ and $\rm H \to ZZ^* \to 4l$ channels, because the
invariant mass resolution due to the  measurement is larger than the
intrinsic width. For Higgs masses
around the WW Boson resonance at 160~GeV$/c^2$, 
the $\rm H\to WW \to ll\nu\nu$ decay is the preferred search channel because the
branching ratio $\rm BR( H\to WW)$  is almost one, but it is not
possible to reconstruct a mass peak because of the two neutrinos. For
masses above the WW resonance, the $\rm H \to ZZ$ channel is again the
most promising search channel.

In the exclusive searches for the Higgs Boson, characteristical
properties of the event topology of the particular production and
decay modes are exploited for event selection. For example, in the
exclusive search for Higgs production through Vector Boson Fusion, the
typical feature are forward jets that are used  to identify
the event. Similarly, in the search for associated Higgs production,
$\rm t\overline{t}H$, the signatures of the two top quarks  are used
for this purpose. In contrast, the inclusive searches do not separate between the various production
topologies. The latter are described in more detail in the following sections.

In case of CMS, the analysis results presented in the following are
based upon publications in the context of the ``Physics Technical
Design Reports''~\cite{CMSPTDR1,CMSPTDR2} (PTDR) published in the year
2006. All analyses apply realistic detector simulations based on
GEANT4, including Level-1 and High-Level Trigger simulations. Where
available, Next-to-Leading Order (NLO) calculations have been used and
systematic errors due to theory and detector effects have been taken
into account.
In the case of ATLAS, the PTDR has been published in 1999 when NLO
calculations and full detector simulations have not been available for the channels discussed
here~\cite{ATLASPTDR1}. The ATLAS collaboration is currently updating the Higgs analyses
according to the most recent simulations and theoretical
calculations. Some  updates on the $\rm H \to \gamma\gamma$
channel are available, but the results on the $\rm H\to ZZ$ and $\rm H
\to WW$ are not official yet and cannot be presented in this paper.

\section{The Channel $\rm H \to \gamma\gamma$}
\label{sec:Hgammagamma} 
The decay into two photons is a rare decay mode with a branching ratio
of 0.2\% at $m_H=120$~GeV$/c^2$~\cite{Djouadi:1997ywMyBib}. The total
NLO cross section times branching ratio, including all production modes is
$\sigma \times \rm B.R. = 99.3$~fb for $m_H = 115$~GeV$/c^2$ and drops
to 41.5~fb for $m_H = 150$~GeV$/c^2$~\cite{Spira:1995mt,CMS_NOTE2006_112}. 
Background processes are separated into reducible and irreducible
backgrounds. Irreducible backgrounds have two real high~$E_t$ photons
produced via born and box diagrams with a cross section of about 80~pb each. Reducible backgrounds arise from
$\gamma$ plus jet or multi-jet events in which one or two jets are
misidentified as photons. The photon identification is
very clean at the ATLAS and CMS detectors. The electromagnetic
calorimeter at ATLAS has a presampler that reduces the fake rate to a
level that is not reached by CMS.  For example, a jet rejection
factor between 4000 and 10000 can be reached at 80\% photon selection
efficiency at ATLAS, depending on 
the transverse momentum~\cite{CarminatiKracow}. This is achieved by
isolation criteria exploiting the fact that misidentified jets are
accompanied by particles measured in the tracker, electromagnetic and
hadronic calorimeters. Therefore, reducible
backgrounds can be suppressed sufficiently. 

\begin{figure}
\begin{center}
% For example, with the option graphicx use
\includegraphics[width=0.49\textwidth]{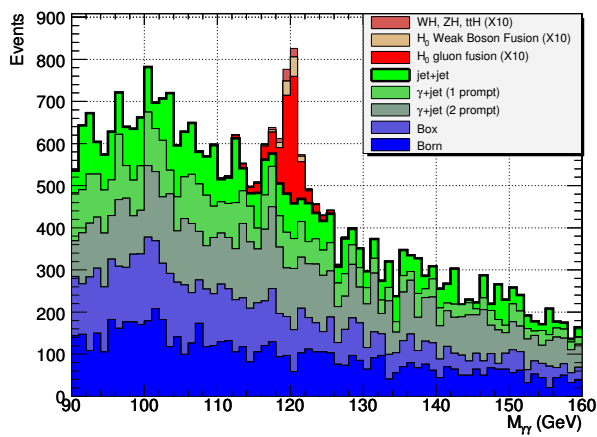}
\caption{Distribution of the invariant mass $m_{\gamma\gamma}$ for
  signal (red, brown) with $m_H = 120$~GeV$/c^2$ and background (blue,
  green) in case of CMS and the neural network analysis. Events are normalized to an integrated luminosity of
  7.7~fb$^{-1}$ and the signal is
  scaled by a factor of 10 for better visibility. \cite{CMS_NOTE2006_112,CMSPTDR2} }
\label{fig:HgammaMass}       % Give a unique label
\end{center}
\end{figure}

Both in ATLAS and CMS the standard cut-based
analyses are supplemented by more powerful separation tools using
neural networks and likelihood methods.
In case of CMS, the cut-based analysis introduces  quality categories
based on the electromagnetic shower shape and pseudo-rapidity. In a
more optimized analysis, a neural network is trained with kinematic
observables in addition to the isolation. These observables are chosen
to be independent of the Higgs Boson mass. The training of the network
is done on the sidebands of the distribution of the invariant mass. This
method can be used for the determination of background rates
directly on data since the narrow peak in the invariant mass
distribution sits on  an almost linear background as illustrated
in Figure~\ref{fig:HgammaMass}.
Systematic errors have a moderate impact on the discovery potential, mostly
because the background can be measured from data. This means that the
error resides in the uncertainty of the fit as well as statistics and the fitting
functions. It has been evaluated to be of the order of 1\%. The error
on the signal is estimated to be about 20\%. About 15\% are
contributed to the theoretical error and the rest to instrumental
effects like luminosity, trigger and tracker material. The error on
the signal affects primarily the determination of exclusion limits
since this has to rely on theoretical predictions. 

The results in terms of observability are  similar for ATLAS and
CMS, even though slightly different methods (neural networks and
likelihoods) and observables have been used. 
 For the cut based analyses, the
discovery significance  is expected to be $6\sigma$ for $m_H = 120$~GeV$/c^2$ 
and for the optimized analyses $10\sigma$ for an integrated
luminosity of $L = 30$~fb$^{-1}$. Details on these
analyses can be found in \cite{CMS_NOTE2006_112,CMSPTDR2,CarminatiKracow,CarminatiActaPhys}.

\section{The Channel $\rm H\to WW \to ll\nu\nu$}
\label{sec:HWW}
For intermediate masses $2m_W < m_H < 2m_Z$ the $\rm H\to WW\to
ll\nu\nu$ channel is expected to be the main discovery channel at
LHC. In this mass range, the $\rm H\to WW$ branching ratio is almost
one. However, no mass peak can be reconstructed because of the two
neutrinos. The normalization of the background is therefore more
difficult. The total NLO signal cross section, including gluon fusion
and vector boson production is largest at $m_H=
160$~GeV$/c^2$ with $\sigma_{NLO} \times \rm{B.R.}(\rm{e,\mu,\tau}) =
2.34$~pb \cite{Davatz:2005pr}. Backgrounds to this channel arise from
continuum di-vector-boson production (WW, ZZ, WZ) with a cross section of
$\sigma_{NLO} \times B.R.(e,\mu,\tau) = 15$~pb. Further backgrounds are
$\rm{t\overline{t}}$ and single top production in association with a W
Boson $\rm tWb$  with  $\sigma_{NLO}
\times B.R.(e,\mu,\tau) = 86.2$~pb and 3.4~pb, respectively. For CMS,
a special technique of re-weighting the $p_t$ spectra of the Higgs
Boson from PYTHIA to the MC@NLO~\cite{mcatnlo,Frixione:2003ei} prediction has been developed and
applied in this analysis~\cite{Davatz:2004zg}. This method of
introducing $p_t$ dependent k-factors has also been used for the WW
background. 

The event selection exploits properties of the event topologies in
order to reject background. For example, the spin correlation between the W
Bosons of the Higgs decay provides a handle to select signal events
based on the angle between the two leptons. Furthermore, the cuts on
missing energy,  the invariant mass of leptons, the transverse momenta
and isolation criteria have been optimized in order to maximize the
discovery significance. In addition, a  central jet veto is applied which
rejects the $\rm{t\overline{t}}$ background by roughly a factor of 30
and signal events only by a factor of about two  \cite{Davatz:2005pr}.

\begin{figure}
\begin{center}
% For example, with the option graphicx use
\includegraphics[width=0.45\textwidth]{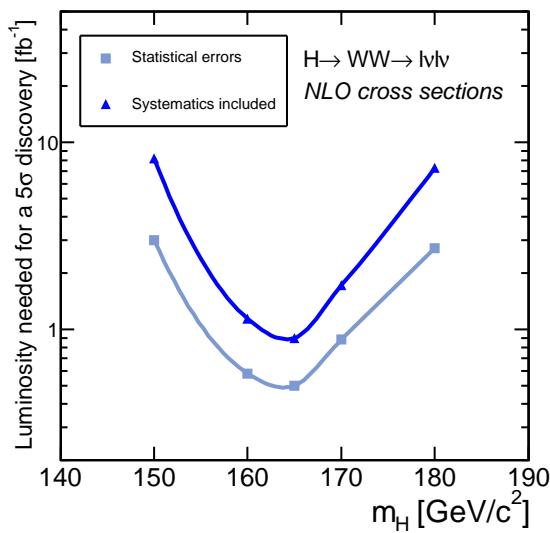}
\caption{Required luminosity for a $5\sigma$ discovery in the  $\rm
  H\to WW \to ll\nu\nu$ channel at CMS. \cite{Davatz:2005pr}}
\label{fig:HWWsigni}       % Give a unique label
\end{center}
\end{figure}

Since this analysis is basically a counting experiment, the
normalization of the background is the largest source of systematic
errors. The $\rm{t\overline{t}}$ can be estimated by replacing the jet
veto with a double b-tag while keeping all other cuts identical. The
expected uncertainty is 16\%. The WZ background can be determined by
requiring a third lepton which gives an uncertainty of 20\%. For the
measurement of the WW background rates, a normalization region in
$\phi_{ll}$ and $m_{ll}$ can be defined, again keeping all other cuts
identical. This results in an expected uncertainty of 17\%. For the
WW background produced by gluon fusion and for the single top background,
it is difficult to define a normalization region and one has to rely
on the theoretical prediction which leads to an uncertainty of
30\%. All these numbers refer to an integrated luminosity of
5~fb$^{-1}$. The resulting effect on the discovery potential in terms
of required luminosity for a $5\sigma$ discovery  is shown in
Figure~\ref{fig:HWWsigni}.

The analysis of this channel is currently being revisited in CMS, in
particular to get better control of the systematics due to missing
transverse energy and jets. Furthermore, an attempt is made to
increase the sensitivity towards lower $m_H$ by e.g. applying
multivariate analysis techniques \cite{bib:HWWimprovements}.

The analysis strategy adopted by ATLAS is similar to the CMS analysis,
but it uses a transverse mass in addition which is defined as $m_T =
\sqrt{2p_t^{ll}E_t^{miss}(1-\cos{\Delta\phi}) }$, where $\Delta\phi$ is the
azimuthal angle between the di-lepton system and the missing transverse energy. This
transverse mass is correlated to the invariant mass of the Higgs Boson and can therefore
be used to define a mass window in order to further reject background
events. In this case a result reaching a significance of $\sigma = 10$
for $m_H = 160$~GeV$/c^2$ 
including a systematic error of 5\% is obtained for an integrated
luminosity of 30~fb$^{-1}$ \cite{ATLASPTDR1}.

\section{The Channel $\rm H\to ZZ \to 4l$}
\label{sec:HZZ}
This channel has a very clean signature due to the presence of four
leptons. It is very promising in the mass range $130$~GeV$/c^2 < m_H <
500$~GeV$/c^2$ except for  $2m_W < m_H < 2m_Z$. The analysis designs
for the different final states (4e, 2e2$\mu$ and 4$\mu$) are very
similar, except for the lepton identification. In the following, the
2e2$\mu$ final state will be described in more detail. 
 The NLO signal cross
section times branching ratio has two maxima, one at $m_H =
150$~GeV$/c^2$ of $\sigma_{NLO} \times \rm{B.R.(2e2\mu)} = 13$~fb, and
another one at  $m_H =
200$~GeV$/c^2$ of $\sigma_{NLO} \times \rm{B.R.(2e2\mu)} = 24$~fb
\cite{HZZNote1}. This behaviour is mostly dominated by the branching
ratio since the cross section itself is continuously falling from 30~pb for
$m_H = 150$~GeV$/c^2$ to 5~pb for $m_H = 500$~GeV$/c^2$. Backgrounds
to this channel are  $\rm{t\overline{t}}$ events with leptonic W Boson
decays and leptons in b-jets which have a cross section of
$\sigma_{NLO} \times \rm{B.R.(2e2\mu)} = 743$~fb. Further backgrounds are Zbb
with $\sigma_{NLO} \times \rm{B.R.(2e2\mu)} = 390$~fb and $\rm ZZ^*/\gamma^*$ events
with $\sigma_{NLO} \times \rm{B.R.(2e2\mu)} = 37$~fb. For the
$\rm ZZ^*/\gamma^*$ background, a re-weighting procedure has been implemented,
which introduces $m_{4l}$-dependent k-factors in order to account for
contributions from all NLO diagrams and from NNLO gluon fusion $\rm
gg\to ZZ^*/\gamma^*$ \cite{HZZNote1}.

\begin{figure}
\begin{center}
\includegraphics[width=0.49\textwidth]{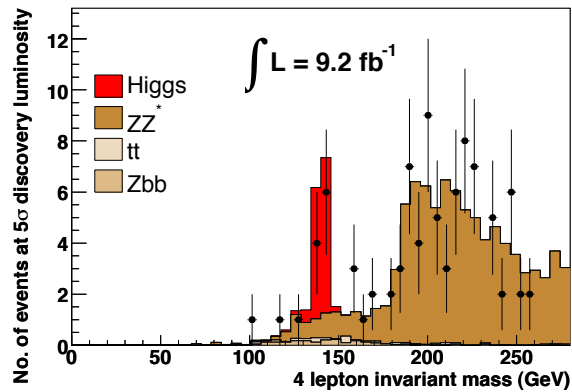}
\caption{Number of expected events for signal and background for an
  integrated luminosity corresponding to a discovery significance of
  $5\sigma$ for a Higgs Boson mass of $m_H = 140$~GeV$/c^2$. As an
  illustration, a toy Monte Carlo distribution based on the histograms is
  superimposed to simulate real CMS data. \cite{HZZNote1} }
\label{fig:HZZMass}       % Give a unique label
\end{center}
\end{figure}

The analysis strategies at CMS and ATLAS are again similar. Both apply
several  tools to reduce the background. Lepton isolation reduces
contributions from leptons in jets. Cuts on the impact parameter of
leptons reduce  b-jets. In addition, leptons are required to
come from the same primary vertex. For lower Higgs Boson masses, one
of the Z Bosons is on-shell, for higher masses with $m_H > 2m_Z$, both
Z Bosons are on-shell.  Mass windows around the Z resonance help to
reduce  $\rm{t\overline{t}}$ and  Zbb backgrounds. By applying these
cuts, the  $\rm{t\overline{t}}$ and  Zbb backgrounds can be suppressed by a
factor of more than 1900 after online selection, while the signal
(with $m_H = 120$~GeV$/c^2$) and
$\rm ZZ^*/\gamma^*$ background are only reduced by a factor of about
two. As an illustration, Figure~\ref{fig:HZZMass} shows the
distribution of the invariant mass of four leptons after offline
selection. 

The systematic uncertainties in this channel are defined by the
uncertainty of the determination of the background rates from
data using sidebands in the mass distribution. The analysis shows that this is possible with a precision of
less than 10\% for Higgs Boson masses below 200~GeV$/c^2$. For higher
masses the uncertainty increases up to 30\% for $m_H =
400$~GeV$/c^2$, because the background is not flat anymore
as visible in Figure~\ref{fig:HZZMass}.

An important alternative to the determination of the background rates
from sidebands is to measure the $\rm Z\to 2l$ process as control
sample and scale it down by a theoretical factor
$\sigma_{ZZ}/\sigma_Z$. This reduces the PDF and QCD scale
uncertainties as well as luminosity uncertainties \cite{CMSPTDR2}. 
The impact of the systematic error on the discovery significance has
found to be small (at the percent level), especially in the low mass range
below 200~GeV$/c^2$.

\section{Summary and Conclusion}
\label{sec:Summary}
\begin{figure}
\begin{center}
\includegraphics[width=0.49\textwidth]{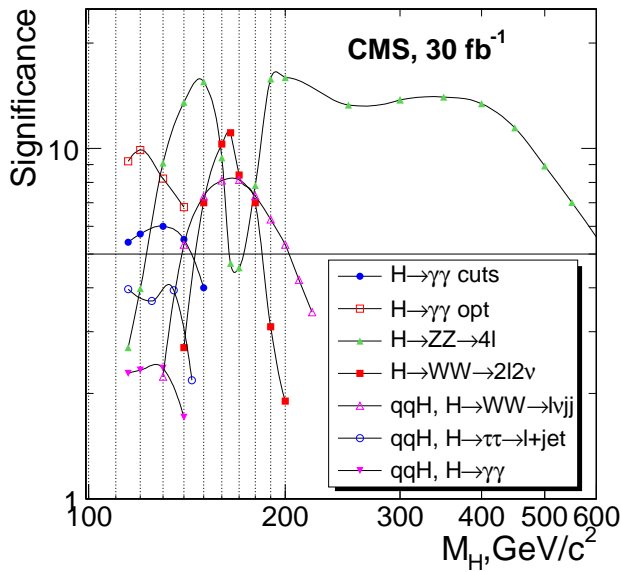}
\caption{Signal significance (in units of $\sigma$) as a function of the Higgs Boson mass for
an integrated luminosity of 30~fb$^{-1}$ at CMS.}
\label{fig:CMSDiscovery}       % Give a unique label
\end{center}
\end{figure}

\begin{figure}
\begin{center}
\includegraphics[width=0.45\textwidth]{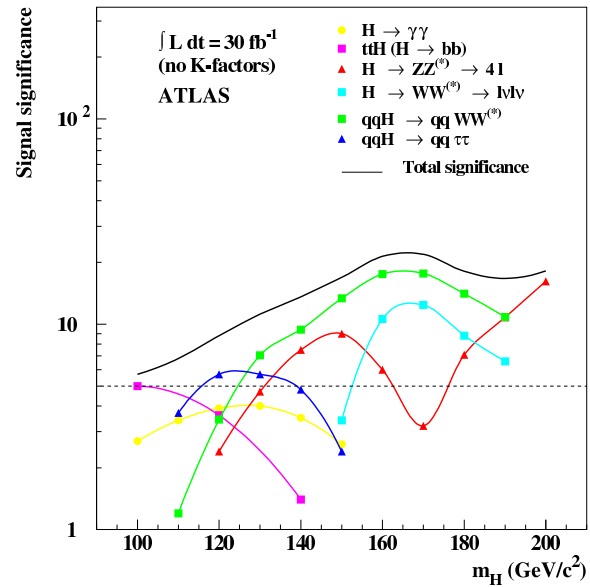}
\caption{Signal significance (in units of $\sigma$) as a function of the Higgs Boson mass for
an integrated luminosity of 30~fb$^{-1}$ at ATLAS. For many channels
the agreement with the CMS results improves if k-factors are introduced. }
\label{fig:ATLASDiscovery}       % Give a unique label
\end{center}
\end{figure}

The three analyses discussed in this paper are complementary in the
sense that they are sensitive to distinct Higgs Boson mass ranges. For
lower masses up to  150~GeV$/c^2$ the $\rm H \to
\gamma\gamma$ channel provides a good discovery potential. For
intermediate masses around 160~GeV$/c^2$ the $\rm H\to WW \to 2l2\nu$ channel is
promising. The $\rm H\to ZZ\to 4l$ channel is interesting for higher
masses, but it also fills a gap at around 140~GeV$/c^2$ where the $\rm
H\to WW$ branching ratio is not yet high enough, and the $\rm H\to
\gamma\gamma$ sensitivity starts to decrease.
By combining all these analyses the full mass range is covered. This is shown in
Figures~\ref{fig:CMSDiscovery}~and~\ref{fig:ATLASDiscovery}. 
From these figures one can conclude that a Standard Model Higgs Boson
is very unlikely to escape the LHC.

\section{Acknowledgements}
\label{sec:Acknowledgements}
I would like to thank Louis Fayard, Sasha Nikitenko, Gigi Rolandi, Markus Schumacher
and Yves Sirois for their valuable suggestions.

\bibliographystyle{ptdr}
\bibliography{myBibSUSY}
%\begin{thebibliography}{999}
%%
%% and use \bibitem to create references.
%Author, \textit{Book title} (Publisher, place year) page numbers
%\end{thebibliography}

\end{document}